
----------
X-Sun-Data-Type: default
X-Sun-Data-Description: default
X-Sun-Data-Name: weyl.tex
X-Sun-Content-Lines: 1400

\input amstex
\magnification 1200
\documentstyle{amsppt}
\NoRunningHeads
\pageheight {7.291in}
\pagewidth {5.5in}
\define\ts{\otimes}
\define\p{\partial}
\define  \kxi{K_{R}\langle x_{i}\rangle}
\define  \kdi{K_{R}\langle \d_{i}\rangle}
\redefine \d{\partial}
\define \krdim{{\text{Kdim}}}
\define \gldim{{\text{gl.dim}}}
\define \gkdim{{\text{GKdim}}}
\redefine \H{{\operatorname {Hom}}}
\define \R{{\Cal R}}
\TagsOnRight
\title  Quantum Weyl Algebras
     \endtitle
\author Anthony Giaquinto and James J. Zhang
\endauthor
\address  Department of Mathematics, University of Michigan, Ann
Arbor MI 48109-1003
    \endaddress
\email tonyg\@math.lsa.umich.edu (A.Giaquinto) and
     jzhang\@math.lsa.umich.edu (J.J. Zhang)
     \endemail
\thanks A. Giaquinto thanks the NSA and J.J. Zhang thanks the NSF
for partial support of this work. Both authors thank
K. Goodearl, R. Irving, S.P. Smith, and J.T. Stafford for useful
conversations about this work.\endthanks


\abstract Let $R$ be a Hecke symmetry. There is then a
natural quantization $A_n(R)$
of the $n^{th}$ Weyl algebra $A_n$ based on $R$. The aim of this paper is to
study some general ring-theoretical aspects of $A_n(R)$ and its relation
to formal deformations of $A_n$.
We also obtain further information on those quantizations obtained from
some well-known Hecke symmetries.

\endabstract
\endtopmatter

\document
\baselineskip 16pt
\head {\bf 0. Introduction} \endhead
Let $K$ be a field and fix an $n$-dimensional vector space $V$.
If $R:V\ts V\rightarrow V\ts V$ is a Hecke symmetry for some $q\in K^*$
then, using the relations given in [WZ], there is a natural quantization
$A_n(R)$ of the $n^{th}$ Weyl algebra $A_n$ based on $R$. This
$A_n(R)$ may be viewed as the algebra of quantized differential
operators on the $R$-symmetric algebra as defined in [Gu]. The
$R$-symmetric algebra is the quantum coordinate
ring of affine $n$-space associated with the quantum function
bialgebra $\Cal O_R(M(n))$ constructed using the method in [FRT].
The aim of most of this paper is to study some ring-theoretical aspects of
$A_n(R)$. Our main
result is that $A_n(R)$ is left and right primitive whenever $q$
is not a root of unity and it is not simple if $\dim _K(A_n(R))=\infty$
and $q\neq \pm 1$. We also show that under some mild assumptions on $R$,
the quantum Weyl algebra $A_n(R)$ is an Auslander regular, Cohen-Macaulay,
Noetherian domain with Gelfand-Kirillov dimension $2n$. Additionally,
we obtain some results on the Krull and global dimensions of those
$A_n(R)$ associated with the standard
multiparameter and
the ``Jordan'' Hecke symmetries. Finally we show that, suitably
interpreted, each $A_n(R)$ is in fact a formal deformation
of $A_n$, and, as such, must be isomorphic to $A_n[[t]]$ as a $K[[t]]$-algebra.
Although we will not use or discuss this fact, we would like
to mention that $A_n(R)$ has been been used to define a quantization
of the universal envoloping algebra $U(\frak {gl}(n))$, see [JZ].

\head {\bf 1. Quantization of linear spaces and Weyl algebras} \endhead
In this section we recall the basic methods of quantizing
linear spaces and differential operators, ({\it {cf}} [FRT], [Gu], and [WZ]).
Fix a field $K$ and let $V$ be an $n$-dimensional vector space
with basis $x_1,\ldots ,x_n$. If $R:V\ts V \rightarrow V\ts V$ is a
linear transformation then $R_{12}=R\ts {\text {Id}}_V$ and
$R_{23}={\text {Id}}_V\ts R$ are
maps $V\ts V\ts V\rightarrow V\ts V\ts V$.
\definition{Definition 1.1}
A linear transformation $R:V\ts V \rightarrow V\ts V$  is a
{\it {Hecke symmetry}} if it satisfies
\roster
\item "(1)" $R_{12}R_{23}R_{12}=R_{23}R_{12}R_{23}$\ -- the
braid relation, and
\item "(2)" $(R-q)(R+q^{-1})=0$ for some $q\in K^*$ -- the Hecke condition.
\endroster \enddefinition
If we view the basis $\{x_i\ts x_j\}$ of $V\ts V$
as an $n^2\times 1$ column vector
lexicographically ordered then $R:V\ts V\rightarrow V\ts V$ may
be represented as an $n^2\times n^2$ matrix (also denoted by $R$) whose
rows and columns are also lexicographically ordered
by pairs $(i,j)$ for $1\leq i,j \leq n$.  Now if
$R(x_i \ts x_j)=\dsize \sum _{k,l}R_{ij}^{kl}x_k\ts x_l$ with
$R_{ij}^{kl}\in K$,
then the entry in row
$(k,l)$ and column $(i,j)$ of $R$ is $R_{ij}^{kl}$.
For some purposes, it will be convenient to present some Hecke symmetries as
an element of the tensor product $M(n)\ts M(n)$,
which may be identified with $M(n^2)$ via
the Kronecker product. Under this identification, the element
$e_{ij}\ts e_{kl}$ of $M(n)\ts M(n)$ corresponds to the
$n^2\times n^2$matrix with entry $1$ in row $(i,k)$ and column $(j,l)$
and zeroes elsewhere.

If $R(x_i\ts x_j)= \dsize \sum _{k,l}R_{ij}^{kl}x_k\ts x_l$ then it
is easy to check that the braid relation (1.1.1) is equivalent to having
$$\sum _{k,l,s}R_{ij}^{kl}R_{lw}^{sv}R_{ks}^{up}=
\sum  _{k,l,s}R_{jw}^{kl}R_{ik}^{us}R_{sl}^{pv}\tag1.2$$
for all $(i,j,w)$ and $(u,p,v)$.

The Hecke condition implies that whenever $q^2\neq -1$ there
is a vector space decomposition
$V\ts V\cong W_+ \oplus W_-$ where $W_+$ and $W_-$ are
eigenspaces for the eigenvalues $q$ and $-q^{-1}$ of $R$
respectively. Specifically, $W_+={\text {Im}}(R-q)$ and
$W_-={\text {Im}}(R+q^{-1})$.
The basic example of a Hecke symmetry is
the permutation operator $\sigma :V\ts V \rightarrow V\ts V$ defined by
$\sigma (x_i\ts x_j)=x_j\ts x_i$ for all $i$ and $j$.  In this case $q=1$ and
$W_+$ and $W_-$ are the subspaces of $V\ts V$ consisting
of the usual symmetric and skew-symmetric tensors
\definition{Definition 1.3}
If  $R:V\ts V \rightarrow V\ts V$ is a Hecke symmetry then the
{\it {$R$-symmetric algebra}}, $K_R\langle x_i\rangle $,
is the quotient $TV/W_+$.\enddefinition
When $K$ has characteristic zero and $q$ is not a root of unity,
a description of $K_R\langle x_i\rangle $ in terms
of a subspace of $TV$ can be found
in [Gu]. To describe
this let
${\text {Sym}}_R(V^{\ts m})=\{\alpha \in V^{\ts m}\, | \, R_i(\alpha)=q\alpha
\quad {\text {for all}} \quad i=1,\ldots , m-1\}$ where $R_i$ acts as $R$ in
tensor factors $i$ and $i+1$ of $V^{\ts m}$ and ${\text {Id}}_V$ in
the others.
There exists a projection operator $P_m:V^{\ts m}\rightarrow
{\text {Sym}}_R(V^{\ts m})$ for each $m$ and so
$\dsize \bigoplus _{m\geq 0}{\text {Sym}}_R(V^{\ts m})$
becomes a $K$-algebra in which the product of $\alpha \in
{\text {Sym}}_R(V^{\ts m_1})$ and $\beta \in
{\text {Sym}}_R(V^{\ts m_2})$ is $P_{m_1+m_2}(\alpha \ts \beta)$.
This approach of viewing $K_R\langle x_i\rangle $
as a subspace of the tensor algebra $TV$ was successfully used to show that,
suitably interpreted, $K_R\langle x_i\rangle $
and the quantum function bialgebra $\Cal O_R(M(n))$ are
formal deformations
of their classical counterparts, {\it cf} [GGS2].

For the purposes of this paper, it will be most convenient
to consider $K_R\langle x_i\rangle $ in terms of generators and
relations. In doing so we have that
$K_R\langle x_i\rangle =K\langle x_i\rangle /I_x$
where $I_x$ is spanned by quadratic relations of the form
$\dsize \sum _{k,l}R_{ij}^{kl}x_kx_l-qx_ix_j$ for all $i$ and $j$.
In the case $R=\sigma$, the relations simply become
$x_i x_j =x_j x_i$ for all $i$ and $j$ and thus
$K_{\sigma}\langle x_i\rangle $ is just the
polynomial ring $K[x_i]$. For many choices of $R$, it is reasonable to
think of $\kxi$ as a graded ``deformation'' of $K[x_i]$ which may be
viewed as
the coordinate ring of
a non-commutative version of affine $n$-space, see Section 4. However, this
is not always the case as there
do exist Hecke symmetries for which $\kxi$ is finite dimensional.

For any $R$-symmetric algebra there is a ``dual'' $R$-symmetric
algebra $K_{R}\langle \d_i\rangle $ obtained from the Hecke
symmetry
$R^*:(V\ts V)^*\rightarrow (V\ts V)^*$. Now $K_{R}\langle \d_i\rangle
=K\langle \d_i \rangle /I_\d$ where $I_\d $ is spanned by
quadratic relations of the form
$\dsize \sum _{k,l}R_{lk}^{ji}\p _k\p _l-q\p _i\p _j$ for
all $i$ and $j$. It is natural to ask whether there is a way to use
$R$ to give an action of the $\d$'s as
``$R$-derivatives'' of $\kxi$ analogous to the
classical case when $R=\sigma$ and each $\d_i$ is the derivation $\d /\d x_i$
of
$K[x_i]$. The appropriate interaction between the $x$'s and
$\d$'s is due to Wess and Zumino and the following can essentially
be found in [WZ].
\definition{Definition 1.4}
The quantum Weyl algebra associated to $R$ is the algebra $A_n(R)$ with
generators $x_1,\ldots,x_n,\p _1,\ldots ,\p _n$ subject, for all
$i$ and $j$ to the relations
\roster
\item "(1)" $\dsize \sum _{k,l}R_{ij}^{kl}x_kx_l=qx_ix_j$.
\item "(2)" $\dsize \sum _{k,l}R_{lk}^{ji}\p _k\p _l=q\p _i\p _j$.
\item "(3)" $\dsize \p _i x_j =\delta _{ij}+q\sum _{k,l}R_{jl}^{ik}x_k \p _l$.
\endroster \enddefinition
We will see that these are natural choices for relations defining
a quantization of $A_n$ based on $R$. Note that
relations (1.4.1) generate $I_x$
and relations (1.4.2) generate $I_\d$ and so there are algebra maps
$K_R\langle x_i\rangle\rightarrow A_n(R)$ and
$K_R\langle \d_i\rangle\rightarrow A_n(R)$. If we set $I_{x,\d}$ to be the
subspace of $K\langle x_i,\d_i\rangle$ generated by relations (1.4.3)
then $A_n(R)=K\langle x_i,\d_i\rangle/(I_x+I_\d+I_{x,\d})$.

Our first step in analyzing the structure of $A_n(R)$ is to
prove the analog of the well-known fact that the monomials
$\{x_1^{i_1}\cdots x_n^{i_n}\d_1^{j_1}\cdots \d_n^{j_n} \}$
form a basis for $A_n$.
\proclaim{Theorem 1.5}
There is a vector space isomorphism $K_R\langle x_i\rangle \ts
K_R\langle \d_i\rangle \cong A_n(R)$.\endproclaim
The proof of this proposition will rely on the following important
technical lemma which guarantees that relations (1.4.3) introduce no
new relations among the $x$'s and $\d$'s other than those given
in (1.4.1) and (1.4.2).
\proclaim{Lemma 1.6} Using relations (1.4.3) we have that, for all $u$
\roster
\item "(1)" $\d_u I_x\subset \dsize \sum _w I_x \d_w$.
\item "(2)" $I_\d x_u\subset \dsize \sum _w x_w I_\d$.
\endroster \endproclaim
\demo{Proof} To prove (1.6.1) it is only necessary to show that
$\p _u(\dsize \sum _{k,l}R_{ij}^{kl}x_kx_l-qx_ix_j)
\subset \dsize \sum _w I_x \d_w$. Now using relations (1.4.3) we get
$$\split
\p _u( \sum _{k,l}R_{ij}^{kl}x_kx_l-qx_ix_j)&=
\sum _{k,l}R_{ij}^{kl}(\delta _{uk}
+q\sum _{p,s}R_{ks}^{up}x_p\p _s)x_l\\&\qquad -
q(\delta _{ui}+q\sum _{p,s}R_{is}^{up}x_p \p _s)x_j\\
&= \sum _{k,l}R_{ij}^{ul}x_l+q\sum _{k,l,p,s}R_{ij}^{kl}R_{ks}^{up}
x_p(\delta _{sl}+q\sum _{v,w}R_{lw}^{sv}x_v \p _w)\\
&\qquad -q\delta _{ui}x_j-q^2 \sum _{p,s}R_{is}^{up}x_p
(\delta _{sj}+q\sum _{v,w}R_{jw}^{sv}x_v \p _w).
\endsplit \tag1.7$$
In order for (1.7) to lie in $\dsize \sum _w I_x\d_w$ we need the sum of all
the
linear terms in the $x$'s to be identically zero since $I_x$ contains
quadratic relations only.
The
linear terms of (1.7) are
$$\sum _{k,l}R_{ij}^{ul}x_l +q\sum _{k,p,s}R_{ij}^{ks}R_{ks}^{up}x_p
-q\delta _{ij}x_j -q^2\sum _p R_{ij}^{up}x_p  $$
which may be written as
$$\sum _{k,l}R_{ij}^{ul}x_l +q\sum _{k,l,s}R_{ij}^{ks}R_{ks}^{ul}x_l
-q\sum _l \delta _{ui} \delta _{lj}x_l -q^2\sum _l R_{ij}^{ul}x_l.\tag1.8$$
Now having (1.8) equal to zero
for all $u$, $i$, and $j$ is the same as having
$$R+qR^2-qR-q^2R=(R-q)(1+qR)=0.$$
Now note that the latter expression is satisfied since $R$
was assumed to satisfy the Hecke condition $(R-q)(R+q^{-1})=0$.

\noindent We now show that the sum of the quadratic terms in the $x$'s of
(1.7) lies in $\dsize \sum _w I_x\d_w$.
These terms are
$$q^2\sum _{k,l,p,s,v,w}R_{ij}^{kl}R_{ks}^{up}R_{lw}^{sv}x_p x_v \p_w
-q^3\sum _{p,s,v,w}R_{is}^{up}R_{jw}^{sv}x_p x_v \p _w.\tag1.9$$
Dividing by $q^2$ and using the braid
relation (1.2) to rewrite the first term of (1.9) we obtain
$$\sum _{k,l,p,s,v,w}R_{ik}^{us}R_{jw}^{kl}R_{sl}^{pv}x_p x_v \p_w
-q\sum _{p,s,v,w}R_{is}^{up}R_{jw}^{sv}x_p x_v \p _w$$
and by reindexing the second sum this becomes
$$\sum _{k,l,p,s,v,w}R_{ik}^{us}R_{jw}^{kl}R_{sl}^{pv}x_p x_v \p_w
-q\sum _{k,l,s,w}R_{ik}^{us}R_{jw}^{kl}x_s x_l \p _w$$
which is an element of $\dsize \sum _w I_x\d_w$.
Similar computations can be used to prove (1.6.2) and the details
are omitted. $\blacksquare$ \enddemo
\demo{Proof of Theorem 1.5} First note that $A_n(R)\cong S/(I_x+I_\d)$ where
$S=K\langle x_i ,\d_i\rangle /I_{x,\d}$. Now using relations (1.4.3) which
generate $I_{x,\d}$ it is
easy to see that every element of $S$ can be
uniquely reduced to the form $\dsize \sum _{I,J}C_{IJ}\, x_I\d_J$ where
$C_{IJ}\in K$ and $x_I$ (resp. $\d_J$) represent the elements
$x_{i_1}\cdots x_{i_r}$ (resp. $\d_{j_1}\cdots \d_{j_s}$) of
$K\langle x_i, \d_i\rangle$.
It follows from the diamond lemma ({\it cf} [B]) that
the elements $\{x_I\d_J\}$ of $S$ are linearly independent. Thus
$\rho : K\langle x_i\rangle\ts K\langle \d_i\rangle
\rightarrow S$ where $\rho (x_I\ts \d_J)=
x_I\d_J$ is a vector space isomorphism. Consequently,
$A_n(R)\cong S/(I_x+I_\d)\cong K\langle x_i\rangle\ts K\langle \d_i\rangle
/(I_x\ts 1+1\ts I_\d)$. By Lemma 1.6, any
element of the two-sided ideal
of $S$ generated by
the relations $I_x\ts 1+1\ts I_\d$ must lie in
$I_x\ts K\langle \d_i\rangle+K\langle x_i\rangle\ts I_\d$ and so
$K\langle x_i\rangle\ts K\langle \d_i\rangle
/(I_x\ts 1+1\ts I_\d)\cong
(K\langle x_i\rangle / I_x)\ts (K\langle \d_i\rangle/I_\d) \cong
K_R\langle x_i\rangle \ts K_R\langle \d_i\rangle$.
$\blacksquare$\enddemo
We will see that the vector space isomorphism
$K_R\langle x_i\rangle \ts
K_R\langle \d_i\rangle \cong A_n(R)$ is crucial for most of the structure
theorems for $A_n(R)$ that are in Section 3. Immediate consequences
of the isomorphism are that
the canonical left and right $A_n(R)$-modules $A_n(R)/(\sum_i A_n(R)\d_i)$
and $A_n(R)/(\sum_i x_iA_n(R))$
may be identified with $ K_R\langle x_i\rangle $ and
$ K_R\langle \d_i\rangle $, respectively.
\head{\bf 2. Some Examples}
\endhead

In this section, we give several examples of quantum Weyl algebras
using the construction of Section 1 for some well-known Hecke
symmetries.
We also state some
nice ring-theoretic properties of these examples; the precise definitions
and proofs of these properties will be given in Section 3.

\noindent {\bf {Example 2.1.}}
This example gives the quantum Weyl algebra associated to the
``standard'' multiparameter $R$-matrix
$$R_{q,p_{ij}}=q\sum_{i}e_{ii}\ts e_{ii}
+\sum_{i<j}(p_{ij}e_{ji}\ts e_{ij}+p_{ij}^{-1}e_{ij}\ts
e_{ji})+(q-q^{-1})\sum_{i>j}e_{ii}\ts e_{jj}.\tag{2.2}$$
which depends on ${n\choose 2} +1$ non-zero scalars $q$ and $p_{ij}$ for $i<j$.
For convenience, we set $p_{ji}=p_{ij}^{-1}$ for $i<j$.
When $n=1$ this is just the $1\times 1$ scalar $q$ and if $n=2$ then
it becomes the $4\times 4$ matrix
$$R_{q,p}=
\pmatrix
q&0&0&0\\0&q-q^{-1}&p^{-1}&0\\0&p&0&0
\\0&0&0&q
\endpmatrix
$$
with $p=p_{12}$.
Note that if $q=p_{ij}=1$ for all $i$ and $j$ then $R_{q,p_{ij}}$ is the
permutation operator $\sigma$.
The quantum groups and quantum linear spaces associated to $R_{q,p_{ij}}$
have been extensively studied in [AST], [GGS2], and [LS]. Some
properties of a multiparameter quantum Weyl algebra
have been studied in [AD] and [J] but there the algebra
differs slightly from $A_n(R_{q,p_{ij}})$;
our primary concern here is with the latter since its relations come
directly from the general construction.
Now, according to Definition 1.4 the
relations of $A_{n}(q,p_{ij})$ are
$$
\align
x_{i}x_{j}&=p_{ij}q x_{j}x_{i},\quad {\text {for all}}\quad i<j\\
\d_{i}\d_{j}&=p_{ij}q^{-1}\d_{j}\d_{i},\quad {\text {for all}}\quad  i<j\\
\d_{i}x_{j}&=p_{ij}^{-1}qx_{j}\d_{i},\quad
{\text {for all}}\quad  i\neq j\tag{2.3}\\
\d_{i}x_{i}&=1+q^{2}x_{i}\d_{i}+(q^{2}-1)\sum_{j>i}x_{j}\d_{j},
\quad {\text {for all}}\quad  i.
\endalign
$$
\noindent
Note when $q=1$ and all $p_{ij}=1$ these relations reduce to those
for $A_n$ and so $A_n(1,1)=A_n$.
The algebra $A_{n}(q,p_{ij})$ is Auslander regular, Cohen-Macaulay
Noetherian
domain with GK dimension $2n$. It is simple if and only if $q^2=1$ and
$char (K)=0$. If $q^2\neq 1$, then $A_{n}(q,p_{ij})$ has Krull
dimension and global dimension $2n$ and
the elements $x_{i}\d_{i}-\d_{i}x_{i}$ are normal (but not invertible)
for all $i$. If $q$ is not root of
1, then $A_{n}(q,p_{ij})$ is left and right primitive.
If $q$ and all $p_{ij}$ are roots of
1, then $A_{n}(q,p_{ij})$ is a PI ring and is not primitive.

If $n=1$, then the resulting quantization of $A_1$ is $A_1(q)$ which has
generators $x \,(=x_1)$ and $\d \,(=\d _1)$ and the single relation
$\d x =1+q^2x\d$. Many of the above results for $A_1(q)$ have been
previously obtained in [G] and [KS].

\noindent {\bf {Example 2.4.}} For $n=2$, there is another
interesting family of quantizations of $A_2$ other than
$A_2(q,p)$ of the previous example. Specifically, let
$$ J_{a,b}=
\pmatrix
1&a&-a&-ab\\0&0&1&b\\0&1&0&-b\\0&0&0&1
\endpmatrix $$
where $a$ and $b$ are two arbitrary scalars in the field $K$.
It follows from [GGS1, 7.1],
that $J_{a,b}$ is a Hecke $R$-matrix with $q=1$. The
relations of
$A_{2}(J_{a,b})$ are
$$\align
x_{1}x_{2}&=x_{2}x_{1}+ax_{1}^{2}\\
\d_{2}\d_{1}&=\d_{1}\d_{2}+b(\d_{2})^{2}\\
\d_{1}x_{1}&=1+x_{1}\d_{1}+ax_{1}\d_{2}\\
\d_{1}x_{2}&=-ax_{1}\d_{1}-abx_{1}\d_{2}+x_{2}\d_{1}+bx_{2}\d_{2}\tag{2.5}\\
\d_{2}x_{1}&=x_{1}\d_{2}\\
\d_{2}x_{2}&=1-bx_{1}\d_{2}+x_{2}\d_{2}
\endalign
$$
If $a=b=0$ then $A_{2}(J_{0,0})$ reduces to the second Weyl algebra $A_{2}$.
Note that in this case
$A_{2}(J_{a,b})\cong A_{2}(J_{\lambda a,\lambda b})$
for any non zero scalar $\lambda \in
K$. The algebra
$A_{2}(J_{a,b})$ is Auslander regular, Cohen-Macaulay Noetherian
domain with GKdimension 4. If $K$ has positive
characteristic, then $A_{2}(J_{a,b})$ is a PI ring
and has Krull
and global dimensions 4. If $K$ has characteristic
zero, then
$A_{2}(J_{a,b})$ is always left and right primitive and it is
simple if and only if
$a=b$.

\noindent {\bf {Example 2.6.}} Our final example illustrates the fact that
$A_n(R)$ may be finite dimensional. Let $n=2$ and consider the
operator
$$\tau=
\pmatrix
-1&0&0&0\\0&0&1&0\\0&1&0&0
\\0&0&0&-1
\endpmatrix .
$$
Note that $\tau $ is a Hecke symmetry with $q=1$. Now $A_2(\tau)$ has relations
$$\align
x_1^2=x_2^2&=\d_1^2=\d_2^2=0\\
x_{1}x_{2}&=x_{2}x_{1}\\
\d_{1}\d_{2}&=\d_{2}\d_{1}\\
\d_{1}x_{1}&=1-x_{1}\d_{1}\\
\d_{1}x_{2}&=x_{2}\d_{1}\tag{2.7}\\
\d_{2}x_{1}&=x_{1}\d_{2}\\
\d_{2}x_{2}&=1-x_{2}\d_{2}
\endalign
$$
and it is easy to check directly that $A_2(\tau)\cong M_2(K)$ (or see
Corollary 3.3),
a simple Artinian
$K$-algebra.

\head{\bf 3. Some Properties of $A_{n}(R)$}
\endhead

In this section we will investigate some ring-theoretical
properties of $A_n(R)$ for an arbitrary Hecke symmetry $R$. Many of these
properties will be proved with the aid of the
element $D=\dsize \sum_{i=1}^nx_i\d_i$ of $A_n(R)$. As an element
of the Weyl algebra $A_n$, the element $D$ is the {\it
Euler derivation}. Namely it is the derivation of $K[x_i]$
sending a homogeneous element $f_m$ of degree $m$ to
$mf_m$. In the quantum case, the action
of $D$ on $\kxi$ will involve the $q$-integers $[m]_{q^{2}}$.
These are defined as follows: set
$[m]_{q^{2}}=\dsize \sum _{i=1}^mq^{2(i-1)}$ for $m\geq 0$ and
$[0]_{q^2}=0$.
Note that
if $q^{2}=1$, then $[m]_{q^{2}}=m$ and if $q^{2}\neq 1$, then $[m]_{q^{2}}=
(q^{2m}-1)/(q^{2}-1)$. The following lemma provides some useful properties
about the element $D$.
\proclaim{Lemma 3.1} (1) Let $f_{m}\in \kxi$ and $g_{m}\in \kdi$
be arbitrary elements of degree $m$, then
$$Df_{m}=[m]_{q^{2}}f_{m}+q^{2m}f_{m}D,\quad {\text{and}}\quad
g_{m}D=[m]_{q^{2}}g_{m}+q^{2m}Dg_{m}.$$
(2) The element $E=1+(q^{2}-1)D$ is a non-zero normal element
in $A_{n}(R)$. Consequently, if $E$ is not invertible, then $A_{n}(R)$
is not simple.

\noindent (3) Let
$$D^{[l]}=\sum_{i_{j}}x_{i_{1}}x_{i_{2}}
\cdots x_{i_{l}}\d_{i_{l}}\cdots \d_{i_{2}}\d_{i_{1}}.$$
Then $D\cdot
D^{[l]}=[l]_{q^{2}}D^{[l]}+q^{2l}D^{[l+1]}$.

\noindent
(4) The (formal) inverse of $E$ is
$\quad E^{-1}=\dsize \sum _{l\geq0}(q^{-2}-1)^{l}D^{[l]}$.
As a consequence, if $\kxi$ or $\kdi$ is finite dimensional,
then $E$ is invertible.

\noindent
(5) For every $l$,
$$D^{[l]}=\prod_{j=0}^{l-1}q^{-2j}(D-[j]_{q^{2}}).$$

\noindent
In particular, if $f_{m}\in \kxi$ and $g_{m}\in \kdi$ are
arbitrary elements of degree $m$, then
$$D^{[l]}f_{m}=\phi(m,l)f_{m}+\sum_{i}s_{i}\d_{i},\quad {\text{and}}\quad
g_{m}D^{[l]}=\phi(m,l)g_{m}+\sum_{i}x_{i}t_{i}$$
for some $s_{i}$ and $t_{i}$ in $A_{n}(R)$, where
$\phi(m,l)=\prod\limits_{j=0}^{l-1}q^{-2j}([m]_{q^{2}}-[j]_{q^{2}})$.
\endproclaim

\demo{Proof} (1) Every polynomial of degree $m$ is a linear
combination of monomials of degree $m$ and so we may assume $f_{m}$ is a
monomial. If $f_{m}=x_{j_{1}}\cdots x_{j_{m}}$ then
$$\align
D\cdot x_{j_{1}}\cdots x_{j_{m}}
&=\sum_{i=1}^{n}x_{i}\d_{i}\cdot x_{j_{1}}\cdots x_{j_{m}}\\
&=\sum_{i}x_{i}\{\delta_{ij_{1}}+q\sum_{k,l}R_{j_{1}l}^{ik}x_{k}
\d_{l}\}\cdot x_{j_{2}}\cdots x_{j_{m}}\\
&=\{x_{j_{1}}+q^{2}x_{j_{1}}\sum x_{l}\d_{l}\}x_{j_{2}}\cdots
x_{j_{m}}\\
&=x_{j_{1}}\cdots x_{j_{n}}+q^{2}x_{j_{1}}Dx_{j_{2}}\cdots x_{j_{m}},
\endalign
$$

\noindent
and, by induction on $m$, we obtain the identity. The second identity is
similar to prove.

\noindent
(2) By (1), it is easy to check that
$Ex_{i}=q^{2}x_{i}E,\quad{\text{and}}\quad \d_{i}E=q^{2}E\d_{i}$
for all $i$ and thus $E$ is normal. By Theorem 1.5 $E\neq 0$.

\noindent
(3) This identity follows easily from part (1).

\noindent (4) Direct computation shows $E\cdot E^{-1}=E^{-1}\cdot E=1$.

\noindent (5) This follows by induction
from the identity in part (3). $\blacksquare$
\enddemo

In terms of the $A_n(R)$-module structure for $\kxi$,
Lemma 4.1.(1) implies that if $f_m$ is a homogeneous element
of degree $m$ in $\kxi$ then $D\cdot f_m=[m]_{q^2}f_m$. In particular,
$D\cdot (f_mf_{m'})=[m+m']_{q^2}f_mf_{m'}=
(D\cdot f_m)f_{m'}+q^{2m}f_m(D\cdot f_{m'})$ and so $D$
is a $\eta$-derivation where $\eta$ is the automorphism of $\kxi$ sending
$f_{m}$ to $q^{2m}f_{m}$.

\proclaim{Theorem 3.2} Suppose that $[m]_{q^{2}}\neq 0$ for all
$m\geq 1$. Then the canonical left module $\kxi$ is faithful and
simple, and ${\text{End}}_{A_{n}(R)}(\kxi)=K$. The same conclusions hold
for the canonical right module $\kdi$, and, as a consequence,
$A_{n}(R)$ is both left and right primitive.
\endproclaim

\demo{Proof} Let $f=\dsize \sum_{l\leq m}f_{l}$ be a non zero element of $\kxi$
where $f_l$ is homogeneous of degree $l$.
If $m=0$, then $f$ is a scalar, so the
submodule generated by $f$ is $\kxi$ itself. If $m>0$,
$$\sum_{i}x_{i}\d_{i}\cdot f=D\cdot \sum_{l}f_{l}=
\sum_{l}[l]_{q^{2}}f_{l}\neq 0 .$$
Hence $\d_{i}f\neq 0$ for some $i$, and the degree of
$\d_{i}f$ is strictly less than the degree of $f$. By induction on
$m$, it follows that the submodule generated by $f$ is must contain a scalar
and
thus $\kxi$ is a simple $A_n(R)$-module.

\noindent To prove $\kxi$ is
faithful, we need to show the annihilator ideal $L$ of the module $\kxi$
is zero. If $g=\dsize \sum C_{IJ}\, x_I\d_J$ let $d(g)$ be the minimum
of $|J|$ appearing in $g$ where $|J|=j_1+\cdots + j_r$ if $J=(j_1,\ldots
j_r)$. Now pick $g\in L$ with minimal $d(g)$, (note that it is necessary
for $d(g)>0$ since $g\cdot 1=0$), and
consider the element
$g'=gD$. By Lemma 4.1.(1), $g'=
\dsize \sum C_{IJ}x_I([|J|]_{q^2}\d _J+q^{2|J|}D\d_J)$. Now since
$d(g')=d(g)$ we must have that $d(gx_i)<d(g)$ for some $i$. But
$g\cdot x_i\in L$ and
this is a
contradiction with minimality
of $d(f)$, so $L=0$.

\noindent Finally, set $f=\theta(1)$ where
$\theta \in {\text{End}}_{A_{n}(R)}(\kxi)$ is arbitrary.
If $f\not\in K$, then $D\cdot f\neq 0$.
But $D\cdot f=D\cdot \theta (1)=\theta (D\cdot 1)=0$
and we obtain a contradiction. Therefore $f\in K$ and
${\text{End}}_{A_{n}(R)}(\kxi)=K$. $\blacksquare$
\enddemo

\proclaim{Corollary 3.3} Suppose that $[m]_{q^2}\neq 0$ for all $m$.

\noindent
(1) $\kxi$ and $\kdi$ have the same Hilbert series.

\noindent
(2) If $\dim(\kxi)=l<\infty$, then $A_{n}(R)\cong M_{l}(K)$, a
simple Artinian algebra of rank $l$.

\noindent
(3) If $\dim(\kxi)=\infty$ and
$q^{2}\neq 1$ then $A_n(R)$ is not simple.
\endproclaim

\demo{Proof} (1) By Theorem 3.2, $\kxi$ is a simple module and so if
$f_{m}\in \kxi $ is a nonzero homogeneous element of degree $m$ there is
some element
$g\in A_{n}(R)$ with $gf_{m}=1$. By counting the degree of $g$, we may
choose $g\in\kdi$ with degree $m$. Hence the dimension of the $m$-homogeneous
component of $\kdi$ is at least the dimension of the $m$-homogeneous
component of $\kxi$. Similarly, the dimension of the $m$-homogeneous
component of $\kxi$ is at least the dimension of the $m$-homogeneous
component of $\kdi$ and therefore these dimensions coincide.

\noindent
(2) By Theorem 3.2, $A_{n}(R)$ is primitive and since its faithful simple
module $\kxi$ is finite dimensional, we have $A_{n}(R)\cong
{\text{End}}_{E_{1}}
(\kxi)$ where $E_{1}={\text{End}}_{A_{n}(R)}(\kxi)=K$. Hence
$A_{n}(R)\cong M_{l}(K)$ where $l=\dim (\kxi)$.

\noindent
(3) According to
Lemma 3.1.(2), we only need to prove $E=1+(q^{2}-1)D$ is not
invertible. It is easy to see that $\phi(m,l)\neq 0$ for $m\gg 0$, which
implies that $D^{[l]}\neq 0$ since $\kxi$ is
infinite dimensional (see Lemma 3.1.(5)). Now by Lemma 3.1.(4), $E^{-1}$ is not
in $A_{n}(R)$
and hence $E$ is not invertible. $\blacksquare$
\enddemo

\noindent
{\bf Remark.} If $\kxi$ and $\kdi$ are finite dimensional then it is
only necessary to require $[m]_{q^{2}}\neq 0$ for all
positive integers $m\leq d$ where $d$ is the maximum degree
of the elements in $\kxi$ and $\kdi$ to obtain the
conclusions of Theorem 3.2 and Corollary 3.3.(1) and (2).

\proclaim{Corollary 3.4} Suppose that $[m]_{q^{2}}=0$ for some $m>1$.
Then $D^{[m]}$ is central and if $A_n(R)$ is a domain, then
$A_n(R)$ is not simple.

\endproclaim

\demo{Proof} By using Lemma 3.1.(1) and (4), it is easy to check
$D^{[m]}$ is central. Now since $A_{n}(R)$ is a domain, $D^{[m]}= 0$ if
and only if $D= [j]_{q^{2}}$ for some $j<m$. But $A_{n}(R)\cong
\kxi\otimes \kdi$ and so
$D$ is not equal to any scalar element in $K$. Thus $D^{[m]}$ is not
zero. Moreover, it is not invertible since $D^{[m]}x_i=0$ in $\kxi$.
Therefore $A_{n}(R)$ is not simple. $\blacksquare$
\enddemo

As a consequence of Corollary 3.3 and Corollary 3.4, if $A_{n}(R)$ is
a simple domain, then $q^2=1$ and $char(K)=0$. The converse, however, is not
true and
we do not know the sufficient conditions on $R$ for $A_n(R)$ to be simple.
However, for those examples with $q^2=1$ discussed in the last section we are
are able to distinguish which are simple.

\proclaim{Theorem 3.5}If char$\,(K)=0$ then $A_n(\pm 1,p_{ij})$ is simple.
\endproclaim
\demo{Proof}
We may assume $q=1$ since $A_n(1,p_{ij})=A_n(-1,-p_{ij})$. Now
according to
Theorem 1.5, it is easy to see that $\{x_1^{i_1}\cdots x_n^{i_n}\d_1^{j_1}
\cdots \d_n^{j_n}\}$ forms a basis for $A_n(1,p_{ij})$. Now $A_n(1,p_{ij})$ is
a
$\Bbb Z^n$-graded algebra with $\deg (x_1^{i_1}\cdots x_n^{i_n}\d_1^{j_1}
\cdots \d_n^{j_n})=(i_1-j_1,\ldots ,i_n-j_n)$. We first show that
$A_n(1,p_{ij})$
is $\Bbb Z^n$-graded simple -- that is, any ideal which contains a non-zero
homogeneous element also contains a scalar and thus must be the entire ring.
Now if $w=\dsize \sum C_{I,J}\,x_1^{i_1}\cdots x_n^{i_n}\d_1^{j_1}
\cdots \d_n^{j_n}$
is homogeneous of degree $(e_1,\ldots ,e_n)$ then direct calculation
shows that for all $m$
$$x_mw-(\prod _{s\neq m}p_{ms}^{e_s})wx_m=
 \dsize \sum j_mC'_{I,J}\,x_1^{i_1}\cdots x_n^{i_n}\d_1^{j_1}
\cdots \d_m ^{j_m-1}\cdots \d_n^{j_n}$$
and
$$\d_mw-(\prod _{s\neq m}p_{ms}^{e_s})w\d_m =
\dsize \sum i_mC''_{I,J}\,x_1^{i_1}\cdots x_m^{i_m-1}\cdots x_n^{i_n}\d_1^{j_1}
\cdots \d_n^{j_n}$$
where $C'_{I,J}$ and $C''_{I,J}$ are non-zero elements of $K$.
By induction, it follows that this ideal contains a scalar.

\noindent Now let $L$ be an arbitrary non-zero ideal of $A_n(1,p_{ij})$ and
suppose
$f\in L$ with $f=f_1+\cdots f_l$ where $f_i$ is homogeneous. Since
$A_n(1,p_{ij})$ is graded simple, we may assume $f_1=1$. For each homogeneous
$a\in
A_n(1,p_{ij})$ the element $af-fa\in L$. By choosing $f$ with minimal $l$, we
get
that each $f_i$ is central.
It follows that $f_i$ must be invertible since $f_i$ is homogeneous and $A$ is
graded simple.
But
the only invertible elements of $A_n(1,p_{ij})$ lie in $K$ and so $f$ is a
scalar
and thus $L=A_n(1,p_{ij})$. $\blacksquare$ \enddemo

Apart from the algebra $A_2(\tau)\cong M_2(K)$ of Example 2.6
which obviously is simple, the other example we have considered
with $q^2=1$ is the algebra $A_2(J_{a,b})$ of Example 2.4. In contrast
with $A_n(1,p_{ij})$ and $A_2(\tau)$, the algebra
$A_2(J_{a,b})$ is not always simple.
\proclaim{Theorem 3.6}If char$\,(K)=0$ then $A_2(J_{a,b})$ is simple if and
only if $a=b$.
\endproclaim
\demo{Proof}
It is easy to see from the defining relations (2.5) for
$A_2(J_{a,b})$ that $\{x_1^m | m\geq 0\}$ is
an Ore set and consequently the localization
$A_2(J_{a,b})[x_1^{-1}]$ is well-defined. The element $x_1^{-1}+(a-b)\d_2$
is normal in $A_2(J_{a,b})$ and is not invertible unless $a=b$. Thus
when $a\neq b$, the localization $A_2(J_{a,b})[x_1^{-1}]$ not simple and
hence neither is $A_2(J_{a,b})$.

\noindent To examine the case when $a=b$ we may, as remarked
earlier, assume that $a=b=1$. To show $A_2(J_{1,1})$ is simple we need
the following lemma which can be found in [W].
\proclaim{Lemma 3.7} Let $T$ be a simple ring, let $\eta$ be an
automorphism of $T$ and let $\delta$ be a $\eta$-derivation. If for
every $m\geq 0$, $D_{m}=\sum_{i=0}^{m}\eta^{i}\delta\eta^{-i}$ is not an
$\eta$-inner derivation, then the Ore extension $T[x;\eta,\delta]$
is simple. $\blacksquare$
\endproclaim

To apply this to our situation note that $A_2(J_{1,1})[x_1^{-1}]$
may be written as an iterated Ore extension
$$A_2(J_{1,1})[x_{1}^{-1}]=K_R\langle
x_{1},x_{2},x_{1}^{-1}\rangle [\d_{2},\eta_{2},
\delta_{2}][\d_{1},\eta_{1},\delta_{1}]$$
where $K_R\langle x_{1},x_{2},x_{2}^{-1}\rangle \cong K\langle
x_{1},x_{2},x_{2}^{-1}\rangle/(x_{1}x_{2}-x_{2}x_{1}-x_{1}^{2})$
and
$$\matrix
\eta_{2}:&x_{1}&\longrightarrow&x_1& \quad &  \delta_{2}:
&x_{1}&\longrightarrow& 0\\
\quad&x_{2}&\longrightarrow&x_{2}-x_1&\quad&\quad
&x_{2}&\longrightarrow&1\\
\eta_{1}:&x_{1}&\longrightarrow&x_{1}&\quad&\delta_{1}:
&x_{1}&\longrightarrow& 1+x_1\d_2\\
\quad&x_{2}&\longrightarrow&x_{2}-x_1&\quad&\quad
&x_{2}&\longrightarrow&(x_2-x_1)\d_2\\
\quad&\d_{2}&\longrightarrow&\d_{2}&\quad&\quad
&\d_{2}&\longrightarrow& -\d_{2}^{2}.
\endmatrix$$
Now it is easy to see that $\eta_2\delta_2=\delta_2\eta_2$ and the
corresponding $D_m$ in Lemma 3.7 is just $(m+1)\delta_2$.
Now we claim that $\delta_2$ is not an $\eta_2$-inner derivation.
Suppose $\eta_2$ is such a derivation, that is, assume that
there is some homogeneous $a\in K_R\langle x_1,x_2,x_1^{-1}\rangle$ with
$\delta_2(f)=\eta_2(f)a-af$ for every
$f\in  K_R\langle x_1,x_2,x_1^{-1}\rangle$.
Now since $0=\delta_2(x_1)=\eta_2(x_1)a-ax_1=x_1a-ax_1$ it follows
that $a$ is a polynomial in $x_1$ and $x_1^{-1}$. Since $\delta_2$ has
degree $-1$ the element $a$ must have the form
$\lambda x_1^{-1}$ for some $\lambda\in K$. But
$1=\delta_2(x_2)=\eta_2(x_2)\cdot \lambda x_1^{-1} -\lambda x_1^{-1}\cdot
x_2=0$ which gives a contradiction and so
$\delta_2$ can not be an $\eta_2$-inner derivation. Now by Lemma 3.7,
$S=K_R\langle x_1,x_2,x_1^{-1}\rangle [\d_2,\eta_2, \delta_2]$ is simple
since $K_R\langle x_1,x_2,x_1^{-1}\rangle\cong A_1[\d^{-1}]$ is simple.
Now we can consider $S[\d_1,\eta_1,\delta_1] \,\, (=A_2(J_{1,1})[x_1^{-1}])$.
In this case, the $D_m$ in Lemma 3.7 satisfies
$D_m(x_1)=(m+1)(1+x_1\d_2)$ and $D_m(\d_2)=-(m+1)\d_2^{2}$.
Now the unique homogeneous solution
to the equation $x_1a-ax_1=(m+1)(1+x_1\d_2)$ is
$(m+1)(x_1^{-2}+x_1^{-1}\d_2)x_2$. For this
choice of $a$, we have that $D_m(\d_2)\neq \eta_1(\d_2)a-a\d_2$ and
therefore $D_m$ is not an $\eta_1$-inner derivation. Once again,
by Lemma 3.7 it follows that $A_2(J_{1,1})[x_1^{-1}]$ is simple.
Finally we need to establish that $A_2(J_{1,1})$ is simple.
Let $I$ be a non-zero ideal of $A_2(J_{1,1})$.
Then $I[x_{1}^{-1}]$ is a non
zero ideal of $A_2(J_{1,1})[x_{1}^{-1}]$
and so $I[x_{1}^{-1}]=A_2(J_{1,1})[x_{1}^{-1}]$
since $A_2(J_{1,1})[x_{1}^{-1}]$ is simple.
Hence $I$ contains $x_{1}^{m}$ for some
$m>0$. But from the defining relations for $A_2(J_{1,1})$ we have that
$\d_1x_1^m-x_1^m\d_1=mx_1^{m-1}+mx_1^m\d_2$. It follows that
$x_1^{m-1}\in I$ since $x_1^m\in I$. By induction on $m$, the ideal
$I$ of $A_2(J_{1,1})$ must contain a scalar and therefore
$I=A_2(J_{1,1})$ and $A_2(J_{1,1})$ is simple.
$\blacksquare$
\enddemo

In the next part of this section we return to study some
general properties of $A_n(R)$ constructed Hecke symmetries $R$
which are ``skew invertible'', by which we mean that
there is a matrix $P$ with
$\dsize \sum _{i,j}P_{gj}^{fi}R_{jl}^{ik}=\delta_{fk}\delta_{gl}=
\dsize \sum _{i,j}R_{gj}^{fi}P_{jl}^{ik}$ for all $f,g,k$ and $l$. This
property on $R$ insures that in $A_n(R)$ every $x_i\d_j$ can be
written as a linear combination of elements of the form $\d_kx_l$.
All of the choices of $R$ used in this paper satisfy this property.

Let $B$
be a
ring and $M$ be a $B$-module. The Krull, global and
Gelfand-Kirillov dimensions of $M$ will be denoted by
$\krdim(M)$ and $\gkdim(M)$ and $\gldim(B)$, respectively.
If $B$ is a Noetherian ring with finite GK
and global dimensions then, $B$ is {\bf Auslander regular} if,
for every Noetherian $B$-module $M$ and every submodule
$N\subseteq {\text{Ext}}^j_B(M,\,B)$, one has
${\text{Ext}}^i_B(N,\,B)=0$ for all $i<j$.
The ring $B$ is {\bf Cohen-Macaulay} if
$j(M)+\gkdim(M)=\gkdim(B)$ holds for every Noetherian $B$-module $M$
where $j(M)=\min\{ j | {\text{Ext}}^j_B(M,\,B)\not=0\}$. The next lemma
contains some facts about these properties which will later be applied
to $A_n(R)$.

\proclaim{Lemma 3.8} (1) Let $B=\dsize \bigcup _{i\geq 0}B_i$ be a filtered
algebra with $B_0=K$. If the associated graded ring $gr(B)$
is an Auslander regular, Cohen-Macaulay, and Noetherian domain with GK
dimension $d$, then $B$ also has these properties.

\noindent
(2) Let $B=\dsize \bigoplus _{i\geq 0}B_i$ be a graded algebra with
$B_0=K$ and suppose that $\eta$ is a graded algebra automorphism and
that $\delta$ is a $\eta$-derivation. If
$B$ is an Auslander regular,
Cohen-Macaulay, and Noetherian domain, then so is the
Ore extension $B[x;\sigma,\delta]$.
\endproclaim
\demo{Proof} See [SZ, 4.4] for (1) and the Lemma of [LS] for (2).
$\blacksquare$ \enddemo

A ring $B$ is an {\bf iterated Ore extension starting
with $K$} if, for each $i=1,\cdots,l$, there is a
subring $B_{i}$ of $B$ with $B_{0}=K$ and $B_{l}=B$ such that
$B_{i}$ is an Ore extension of $B_{i-1}$.

\proclaim{Lemma 3.9} Suppose $R$ is a skew-invertible
Hecke $R$-matrix such that
$\kxi$ and $\kdi$ are iterated Ore extensions starting with $K$.
If the relations between $x_{1},\cdots, x_{n},\d_{1},\cdots,\d_{l}$
do not involve $\d_{l+1},\cdots, \d_{n}$  then
$A_{n}(R)$ is an iterated Ore extensions starting with $K$.
\endproclaim

\demo{Proof} Let $B_{l}$ be the subring of $A_n(R)$ generated by
$x_{1},\cdots, x_{n},\d_{1},\cdots,\d_{l}$.
Clearly, $B_0=\kxi$ and $B_n=A_n(R)$.
We must show that
$B_l$ is an Ore extension of $B_{l-1}$.
First note that Theorem 1.5 and the
fact that $\kxi$ and $\kdi$ are iterated Ore extensions starting with $K$
imply that the
set of monomials
$$\{x_{1}^{i_{1}}\cdots x_{n}^{i_{n}}\d_{1}^{j_{1}}\cdots
\d_{l}^{j_{l}}\, |\, i_{s},j_{t}\geq 0\}$$
is linearly independent. Now since the relations between $x_{1},\cdots,
x_{n}$ and $\d_{1},\cdots,\d_{l}$ do not involve $\d_{l+1},
\cdots, \d_{n}$, the above set must span $B_{l}$.
Thus $B_{l}=B_{l-1}[\d_{l};\eta,\delta]$ for some ring
endomorphism $\eta$ of $B_{l-1}$ and $\eta$-derivation $\delta$.
Since $R$ is skew-invertible, the endomorphism $\eta$ must actually be an
automorphism. $\blacksquare$
\enddemo

\noindent {\bf Remarks:} (1) The conclusion of this lemma clearly holds when
$\kxi$ and $\kdi$ are iterated Ore extensions and the relations
between $x_1,\ldots, x_n, \d_{l},\ldots \d_n$ do not involve
$\d_1,\ldots \d_{l-1}$. Note that this is the case for the
quantizations $A_n(q,p_{ij})$ and $A_2(J_{a,b})$.

\noindent (2) In the same way, it is easy to prove that $gr(A_n(R))$ is an
iterated
Ore extension starting with $K$.

As a consequence of Lemmas 3.8 and 3.9, we have the following:

\proclaim{Corollary 3.10} Let $R$ satisfy the same conditions as in Lemma 3.9.

\noindent
(1) $A_{n}(R)$ and $gr(A_{n}(R))$ are Auslander regular,
Cohen-Macaulay, and Noetherian domains.

\noindent
(2) The global and GK dimensions of $gr(A_{n}(R))$ are both $2n$.

\noindent
(3) The Krull and global dimensions of $A_n(R)$ are both $\leq 2n$ and
its $GK$-dimension is $2n$.
\endproclaim

\demo{Proof} (1) This follows immediately from Lemma 3.8 and Lemma 3.9.

\noindent
(2) Since any connected graded ring has the simple module $K$, we obtain
from [MR, 7.9.18] that
the global dimension of the graded Ore extension
$gr(A_{n}(R))$ is $2n$.
By Theorem 1.5, the GK dimensions of $gr(A_{n}(R))$ and $A_n(R)$ are both $2n$.

\noindent
(3) The bounds on the Krull and global dimensions follow from
[MR, 6.5.4 and 7.5.3]. $\blacksquare$
\enddemo

Next we apply the previous results to further study some of the examples
described in Section 2.

\proclaim{Theorem 3.11}(1) The algebra $A_{n}(q,p_{ij})$ is an
Auslander regular, Cohen-Macaulay, Noetherian domain with GK
dimension $2n$.

\noindent (2) $A_{n}(q,p_{ij})$ is a PI ring if and only if
$[m]_{q^{2}}=0$ and
$p_{ij}^{m}=1$ for some $m$ and all $i$ and $j$.

\noindent (3) If $q^2\neq  1$ or $char(K)\neq 0$,
then $A_{n}(q,p_{ij})$ has Krull dimension and global dimension $2n$.
\endproclaim

\demo{Proof} (1) This follows from Corollary 3.10 since it is well known that
for this example both $\kxi$ and
$\kdi$ are iterated Ore extensions starting with $K$,
and the relations between
$x_{1},\cdots, x_{n}$, $\d_{l+1},\cdots,\d_{n}$ are independent of
$\d_{1},\cdots, \d_{l}$ for all $l$.

\noindent (2) If $A_{n}(q,p_{ij})$ is a PI ring, then the subring generated by
$x_i$ and $x_j$ is also a PI ring and hence some power of $p_{ij}q$
is one since $x_ix_j=p_{ij}qx_jx_i$.
Similarly, we get that some power of $p_{ij}^{-1}q$ is 1
by considering the subalgebra generated by $\d_i$ and $\d_j$.
Consequently, there exist an integer $m$ such that
$p_{ij}^m=1$ and $q^{2m}=1$ for all $i$ and $j$.
Now if $q^2\neq 1$, then we also have $[m]_{q^{2}}=0$.
If, however, $q=\pm 1$ then we must have that $char(K)=p>0$ since otherwise
$A_n(q,p_{ij})$ would be simple by Theorem 3.5. Now $m'=pm$ satisfies
$[m']_{q^2}=0$
and $p_{ij}^{m'}=1$ for all $i$ and $j$.
Conversely, if $[m]_{q^{2}}=0$ and
$p_{ij}^{m}=1$ for some $m$ then $x_{i}^{m}$ and $\d_{i}^{m}$
are central for all $i$, and thus
$A_{n}(q,p_{ij})$ is a PI ring.

\noindent (3)
We will prove the results on the Krull and global dimensions with the
aid of a suitable regular sequence. Recall that elements
$a_1,\cdots ,a_l$ of a ring $T$ form a regular sequence if
$a_{i+1}$ is a regular normal element in $T_i=T/(a_1,\ldots ,a_i)$ for
each $i=0,\ldots l-1$.

First assume that $q^2\neq 1$. From the defining relations of
$A_{n}(q,p_{ij})$ it is easy to check that
$1+(q^{2}-1)x_{n}\d_{n}$, $x_{n-1}$, $\d_{n-1}$, $\cdots$, $x_{1}$,
$\d_{1}$, $x_{n}-1$ is a regular sequence of $A_{n}(q,p_{ij})$.
Now from [MR,6.3.9] and  [MR, 7.3.5]
we obtain
$\krdim(A_{n}(q,p_{ij}))\geq 2n$ and
$\gldim(A_{n}(q,p_{ij}))\geq 2n$. Together with Corollary 3.10.(3), this
implies that the
Krull and global dimensions of
$A_{n}(q,p_{ij})$ are both $2n$.

If $char(K)=p>0$ then we may assume $q^2=1$.
Now since $A_{n}(q,p_{ij})=A_{n}(-q,-p_{ij})$ we may further assume
that $q=1$. In this case it is again routine to check that
$x_{1}^{p}$, $\d_{1}^{p}$, $\cdots$, $x_{n}^{p}$, $\d_{n}^{p}$ forms a sequence
of regular elements in $A_{n}(1,p_{ij})$.
Moreover, the
algebra $A_{n}(1,p_{ij})/(x_{i}^{p},\d_{i}^{p})$ is finite
dimensional and so by [MR, 6.3.9] and Corollary 3.10.(3) we have
$\krdim(A_{n}(1,p_{ij}))=2n$.
We can find the global dimension from the Cohen-Macaulay property.
Since $A_{n}(1,p_{ij})/(x_{i}^{p},\d_{i}^{p})$ is finite
dimensional, $\gkdim(A_{n}(1,p_{ij})/(x_{i}^{p},\d_{i}^{p}))=0$, and
$j(A_{n}(1,p_{ij})/(x_{i}^{p},\d_{i}^{p}))=\gkdim(A_{n}(1,p_{ij}))-0=2n$.
Thus
$\gldim(A_{n}(1,p_{ij}))\geq
j((A_{n}(1,p_{ij})/(x_{i}^{p},\d_{i}^{p}))=2n$ and then Corollarry 3.10.(3)
forces $\gldim(A_{n}(1,p_{ij}))=2n$. $\blacksquare$
\enddemo

\noindent {\bf {Remark 3.12}} (1) In a similar fashion, we can prove that the
Krull and global
dimensions are $2n$ for the quantum Weyl algebra $A_{n}^{\bar{q},\Lambda}$
studied in [AD] and [J].

\noindent (2) If $char(K)=0$ then S.P. Smith has shown that
$\krdim (A_n(1, p_{ij}))=n$. We conjecture that $\gldim (A_n(1, p_{ij}))=n$
as well.

We now turn to further study the algebra $A_2(J_{a,b})$.
First we start with the following
elementary result.

\proclaim{Proposition 3.13} The algebra $A_2(J_{a,b})$ is an Auslander regular,
Cohen-Macaulay, Noetherian domain with GK dimension 4.
\endproclaim

\demo{Proof} First note that $J_{a,b}$ is skew-invertible and both
$K_{J_{a,b}}\langle x_1, x_2\rangle$ and
$K_{J_{a,b}}\langle \d_1, \d_2\rangle$ are iterated Ore extensions
starting with $K$. Now the results follow from Corollary 3.10 since the
relations between $x_1$, $x_2$ and $\d_2$ do not involve $\d_1$. $\blacksquare$
\enddemo
\proclaim{Lemma 3.13} Let $K$ be a field of characteristic $p>0$ and let
$R$ be an affine PI $K$-algebra which is a finite module over its
Noetherian center $C$. If
$\sigma$ is an automorphism of $R$ of finite order and $\delta$ is a
$\sigma$-derivation, then the Ore extension $R[x;\sigma,\delta]$ is also a
PI ring.
\endproclaim

\demo{Proof} By the Artin-Tate Lemma [MR, 13.9.10], $C$
is affine. For every central element $c$, $\sigma(c)\in C$ and so $\sigma$
restricts to an automorphism of $C$.
Since $\sigma$ has finite
order, $C$ must be integral over the fixed subring $C^{\sigma}=\{c\in C \,|\,
\sigma(c)=c\}$ and thus $C$ is a finitely generated
$C^{\sigma}$-module since $C$ is affine. Hence $C^{\sigma}$ is also affine by
the Artin-Tate Lemma
[MR, 13.9.10]. Now for every $r\in C^\sigma$, we have
$$xr=rx+\delta(r)$$
where $\delta(r)\in R$. Consider the subring $C'$ generated by
the set $\{r^{p}| {\text { for all }} r\in C^\sigma\}$.
If the algebra $C^\sigma$ is generated by $\{c_1,\cdots, c_t\}$,
then  $C'$ is generated by $\{ c_1^p,\cdots, c_t^p\}$. Consequently,
$C'$ is affine and $C^\sigma$ is finitely generated over $C'$. For
every $r=c^p\in C'$, it follows that
$$x r=xc^p=c^p x+pc^{p-1}\delta(c)=c^p x=rx\tag{3.14}$$
and hence (3.14) holds for all $r\in C'$. This implies that the subalgebra
of $R[x;\sigma,\delta]$ generated by $C'$ and $x$ is isomorphic to the
commutative algebra $C'[x]$. Since $R$ is a finite
$C'$-module, the Ore extension $R[x;\sigma,\delta]$ must be a finite
$C'[x]$-module. From [MR, 13.4.9], we conclude that $R[x;\sigma,\delta]$ is a
PI
algebra. $\blacksquare$
\enddemo

\proclaim{Theorem 3.15} The algebra $A_2(J_{a,b})$ is a PI ring if and only if
$char(K)=p>0$. In this case, $A_2(J_{a,b})$ is a finite module over its
Noetherian center.
\endproclaim

\demo{Proof} If $char(K)=0$, Theorem 3.2 implies that $A_2(J_{a,b})$ is not a
PI ring. Now assume $char(K)=p>0$. By Lemma 3.13, $K_{R}\langle
x_1,x_2\rangle=K[x_1][x_2; 1, \delta]$ is  a PI ring and hence
$K_{R}\langle x_1,x_2\rangle$ is finite module over its Noetherian
center, see [St, 2.12].
Similar to the proof of Theorem 3.6, it is easy to see that
$$A_2(J_{a,b})=K_{R}\langle x_1,x_2\rangle[\partial_2;
\sigma_2,\delta_2] [\partial_1; \sigma_1,\delta_1]$$
where
$$\sigma_2: x_1 \longrightarrow  x_1, \;\; x_2\longrightarrow
x_2-bx_1, \;\;\;{\text{and}}$$
$$\sigma_1: x_1 \longrightarrow  x_1, \; x_2\longrightarrow
x_2-ax_1, \partial_2 \longrightarrow \partial_2.$$
It is easy to check that $\sigma_1$ and $\sigma_2$ have finite order
$p$. Then the assertions of the theorem follow by applying Lemma 3.13 and [St,
2.12].
Note that these rings satisfy the conditions
in [St, 2.12] by Lemma 3.9, Corollary 3.10, and Quillen's Theorem
(see [MR, 12.6.13]). $\blacksquare$
\enddemo

\proclaim{Corollary 3.15} If $char(K)=p>0$ then the Krull
and global dimensions of $A_2(J_{a,b})$ are both 4. \endproclaim

\demo{Proof} Since $A_2(J_{a,b})$ is a semiprime
Noetherian PI ring, it follows
from [MR, 6.4.8 and 10.10.6] that
$\krdim(A_2(J_{a,b}))=4$. From [RSS], we get
$\krdim(A_2(J_{a,b}))\leq \gldim(A_2(J_{a,b}))$. But by Corollary 3.10.(3)
$\gldim(A_2(J_{a,b}))\leq 4$ and finally we obtain
$\gldim(A_2(J_{a,b}))=4$. $\blacksquare$
\enddemo
\head{\bf 4. Deformations and cohomology of Weyl algebras}
\endhead

In this section we discuss the formal deformation theory of
the Weyl algebra $A_n$ and its relation to the quantum Weyl algebras
$A_n(R)$. In particular, we will see that, suitably interpreted,
$A_n(R)$ is a deformation of the classical Weyl algebra
$A_n$. Throughout this section assume $char(K)=0$.
We first state a result from [Sr] concerning
the Hochschild cohomology groups
$H^\ast (A_n,A_n)$. This will be useful in determining
which types of deformations
of $A_n$ that are possible.

\proclaim{Theorem 4.1}[Sr] The Hochschild cohomology groups
$H^m(A_n,A_n)=0$ for all $m>0$. $\blacksquare$ \endproclaim

We now turn to the deformation theory of the Weyl algebras.
First we briefly recall the basic
definitions and notions of algebraic deformation theory,
{\it cf} [Ge].
\definition{Definition 4.2}
A $K[[t]]$-algebra $A_t$ is a {\it formal deformation}
of a $K$-algebra $A$ if it is a flat, $t$-adically complete
$K[[t]]$-module equipped with a $K$-algebra isomorphism
$A_t\ts_{K[[t]]}K\cong A$.\enddefinition
Since $K$ is a field, the flatness hypothesis of the
definition can simply be replaced with $t$-torsion free;
(the former concept is required when considering algebras
over a commutative ring).
When $A_t$ is a deformation of $A$ we may identify
$A_t$ with $A[[t]]$ as $K[[t]]$-modules and, in doing
so, the deformed multiplication
$\mu _t:A[[t]]\ts _{K[[t]]}A[[t]]\rightarrow A[[t]]$
necessarily has the
form $\mu_t=\mu +\mu_1t+\mu_2t^2+\cdots $ where $\mu$ is the
multiplication in $A$ and each $\mu _i\in \H _K (A\ts A,A) $
is extended to be $K[[t]]$-linear. The {\it trivial deformation} has
$\mu_i=0$ for all $i\geq 1$ and so is
the algebra $A[[t]]$.
Deformations $A_t$ and
$A_t'$ are {\it equivalent} if there is a $K[[t]]$-algebra isomorphism
$\phi _t:A_t\rightarrow A_t'$ of the form
$\phi _t={\text {Id}}_A+\phi _1t +\phi _2t^2 +\cdots $ where each
$\phi _i\in \H (A,A)$ is extended to be $K[[t]]$-linear.
An algebra is {\it rigid} if every deformation is equivalent to the
trivial deformation.

As shown in [Ge], the associativity condition $\mu_t((\mu_t(a,b),c)=
\mu_t((a,\mu_t(b,c))$ imposes restrictions on the $\mu_i$ which are
intimately associated with the Hochshild cohomology groups
$H^*(A,A)$ in low dimensions. We shall not need or describe
this connection in detail, but we note in particular that the first
non-zero $\mu_r$ is a Hochschild 2-cocycle and can be changed by any coboundary
by passing to an equivalent deformation. Thus, if $\mu_r$ is itself
a coboundary, then there is an equivalent deformation with
$\mu_i=0$ for all $i\leq r$. Consequently, $A$ is rigid
whenever $H^2(A,A)=0$; (the converse is false if $K$ has positive
characteristic and it is unknown if $K$ has
characteristic 0).

In order to study
the relation of the quantizations $A_n(R)$ to deformations of $A_n$ we must
first make some modifications in their definition since, as defined
in Section 1, $A_n(R)$ is a $K$-algebra while deformations are
$K[[t]]$-algebras.
Throughout the remainder of this section let $V_t=V[[t]]=V\ts K[[t]]$ and
view all tensor products over $K[[t]]$ ({\it i.e.} $\ts =\ts _K[[t]]$).
Let $\R:V_t\ts V_t\rightarrow V_t \ts V_t$ be a
linear transformation
of the form $\R=\sigma +t\R_1+t^2\R_2+\cdots $ where $\sigma$ is again the
operator sending $a\ts b$ to $b\ts a$ for $a$ and $b$ in $V_t$.
In this context,
we say $\R$ is a formal Hecke symmetry if it satisfies the braid
relation, $\R_{12}\R_{23}\R_{12}=\R_{23}\R_{12}\R_{23}$ as linear
maps $V_t\ts V_t \ts V_t\rightarrow V_t\ts V_t \ts V_t$, and the
formal Hecke condition $(\R-q(t))(\R+q(t)^{-1})=0$ for some
$q(t)\in K[[t]]$ with $q(0)=1$.
Now if $\R(x_i\ts x_j)=\dsize \sum _{k,l}\R_{ij}^{kl}
x_k\ts x_l$ with $\R_{ij}^{kl}\in K[[t]]$, we define $A_n(\R)$ to be
the quotient of $K\langle x_1,\cdots x_n,\d_1,\cdots \d_n\rangle [[t]]$
subject to the relations
\roster
\item "(a)" $\dsize \sum _{k,l}\R_{ij}^{kl}x_kx_l=q(t)x_ix_j$.
\item "(b)" $\dsize \sum _{k,l}\R_{lk}^{ji}\d _k \d_l=
q(t)\d _i\d _j.${\hfil {(4.3)}}
\item "(c)" $\dsize \d _i x_j =
\delta _{ij}+q(t)\sum _{k,l} \R_{jl}^{ik}x_k \d _l$.
\endroster
These relations are of course obtained
from those for $A_n(R)$ by replacing $R_{ij}^{kl}$ with $\R_{ij}^{kl}$.
A natural question is to determine whether $A_n(\R)$ is a deformation
of $A_n$. By definition, $A_n(\R)$ is $t$-adically complete and,
moreover, $A_n(\R)\ts _{K[[t]]}K\cong A_n$ since the defining relations
of $A_n(\R)$ reduce to those for $A_n$ when $t=0$. Thus $A_n(\R)$ will be a
deformation
of $A_n$ if and only if it is $t$-torsion free. Now since $\R$ is a
formal Hecke symmetry, the results of [GGS2] imply that $K_{\R}
\langle x_i\rangle$
and $K_{\R}\langle \d _i\rangle$ are deformations of the polynomial
rings $K[x_i]$ and $K[\d_i]$ and consequently
there are $K[[t]]$-module isomorphisms $K_{\R}
\langle x_i\rangle\cong K[x_i][[t]]$ and $K_{\R}
\langle \d_i\rangle\cong K[\d_i][[t]]$. Now reasoning in the same
way as in Theorem 1.5, there is a $K[[t]]$-module isomorphism
$A_n(\R)\cong K_{\R}\langle x_i\rangle \widehat {\ts}
K_{\R}\langle \d_i\rangle$ where $\widehat \ts$ indicates the completion
of the algebraic tensor produce $\ts _{K[[t]]}$ with respect to the
$t$-adic topology. Combining these facts gives a $K[[t]]$-module
isomorphism $A_n(\R)\cong (K[x_i] \ts K[\d_i])[[t]]$ and the latter module
is isomorphic to $A_n[[t]]$. Thus we have the following:
\proclaim{Theorem 4.4}If $\R$ is a formal Hecke symmetry then
$A_n(\R)$ is a deformation of $A_n$.
 $\blacksquare$\endproclaim
Now the Weyl algebra $A_n$ is rigid
since, according to Theorem 4,1,
$H^2(A_n,A_n)=0$. Thus for any formal Hecke symmetry there is an
algebra isomorphism $A_n(\R)\cong A_n[[t]]$.
As an illustration of this fact, consider the quantization
multiparameter quantization $A_n(R_{q,p_{ij}})$ of Example 2.1.
Now if we replace $q$ with $e^ t$ and $p_{ij}$ with $e^{c_{ij}t}$
for $c_{ij}\in K$ in the definition of $R_{q,p_{ij}}$ then
we obtain a formal Hecke symmetry $\R_{q,p_{ij}}$. Consequently,
$A_n(\R_{q,p_{ij}})$ is a deformation of $A_n$ and
$A_n(\R_{q,p_{ij}})\cong A_n[[t]]$. For the classical case when all
$p_{ij}=1$ this isomorphism has been explicitely
constructed in [Og]. It is important to note that, even if a
Hecke symmetry $R$ is obtainable as a specialization of a formal
Hecke symmetry $\R$, the
rigidity of $A_n$ does not guarantee the existence of
a $K$-algebra isomorphism $A_n(R)\cong A_n$. This may be seen even in the
``simplest'' quantum Weyl algebra, $A_1(q)$. If $q\neq 1$
then $A_1(q)$ is not simple and thus is not isomorphic to $A_1$.
To further illustrate this point, note that
in $A_1[[t]]$ the elements $x$ and $\d_t=\dsize \frac
{1}{x}\cdot \frac {e^{tx\d}-1}{e^t-1}$ satisfy
$\d _tx=1+q^2x\d_t$ with $q=e^{t/2}$ and thus provide an explicit
isomorphism $A_1[[t]]\cong A_1(e^{t/2})$. Even if $K=\Bbb C$ this analytic
isomorphism has zero radius of convergence.

\enddocument